\documentclass[twocolumn,superscriptaddress,longbibliography,aps,preprintnumbers]{revtex4-2}

\usepackage{graphicx,color}
\usepackage{xcolor}

\usepackage{bm,bbm,amsmath,amssymb,braket}
\usepackage{epstopdf}
\usepackage{relsize}
\usepackage{epstopdf}
\usepackage{comment}
\usepackage{soul}
\usepackage{float}
\usepackage{lipsum}
\usepackage[urlcolor=blue,colorlinks=true,citecolor=blue,
	linkcolor=red,pdfstartview={FitH},bookmarks=false]{hyperref}
\usepackage{MnSymbol}
\usepackage{stmaryrd}
\usepackage{mathtools}

\graphicspath{{Figures/}{./Figures/}{.}{./paper_figures/}}
\setstcolor{red}

\definecolor{purple}{rgb}{0.5, 0., 0.8}

\definecolor{persianblue}{rgb}{0.11, 0.22, 0.73}

\definecolor{darkred}{RGB}{182, 24, 24}

\usepackage{amssymb}

\begin{document}
\title{Arrival-time problem in the quantum information era -- observation of a Zeno crossover in monitored arrival-time distributions}

\author{Yousef Mafi}\email{Email: yousef.mafi@tuni.fi}
\affiliation{Computational Physics Laboratory, Physics Unit, Faculty of Engineering and Natural Sciences, Tampere University, P.O. Box 692, FI-33014 Tampere, Finland}

\author{K. G. S. H. Gunawardana}
\affiliation{Computational Physics Laboratory, Physics Unit, Faculty of Engineering and Natural Sciences, Tampere University, P.O. Box 692, FI-33014 Tampere, Finland}

\author{Kim P\"oyh\"onen}
\affiliation{Computational Physics Laboratory, Physics Unit, Faculty of Engineering and Natural Sciences, Tampere University, P.O. Box 692, FI-33014 Tampere, Finland}

\author{Teemu Ojanen}\email{Email: teemu.ojanen@tuni.fi}
\affiliation{Computational Physics Laboratory, Physics Unit, Faculty of Engineering and Natural Sciences, Tampere University, P.O. Box 692, FI-33014 Tampere, Finland}

\date{\today}

\begin{abstract}
When does a quantum particle arrive at a detector? The temporal distributions of itinerant quantum particles have been debated since the birth of quantum mechanics. Here we study the arrival-time problem in periodically monitored spin chains and quantum circuits. The arrival dynamics exhibits two regimes, the quantum Zeno and the low-monitoring regime, separated by an optimal monitoring period set by the single-particle bandwidth. In support of the theory, on superconducting quantum hardware, we measure arrival-time distributions for single and multiple interacting particles in one- and two-dimensional circuits. The measured distributions resolve both regimes and agree closely with quantum-trajectory theory. The crossover persists in interacting many-particle systems: the optimal monitoring point is preserved while many-body interference reshapes the distributions. Zeno suppression likewise persists in two-dimensional geometry. Together, these results show that the two regimes are a generic feature of monitored propagation rather than a single-particle effect. To our knowledge, our work constitutes the first experimental measurement of arrival-time distributions in a monitored quantum transport setting, addressing one of the longest-standing problems in quantum mechanics.
\end{abstract}

\maketitle

\section{Introduction}

The wave-like nature of matter, first spectacularly confirmed in 1927 through
the electron diffraction experiments of Davisson and Germer and of Thomson~\cite{DavissonGermer1927,ThomsonReid1927}, manifests through striking spatial patterns in particle distributions. While predicting spatial
distributions of itinerant particles has become a standard textbook application of the Born rule, the quantum time-of-arrival problem remains
debated: there is still no universally accepted arrival-time observable or probability distribution for detector click times~\cite{Allcock1969,Kijowski1974,Aharonov1998,MugaLeavens2000,Roncallo2023}.

\begin{figure}[ht!]
    \centering
    \includegraphics[width=1\columnwidth]{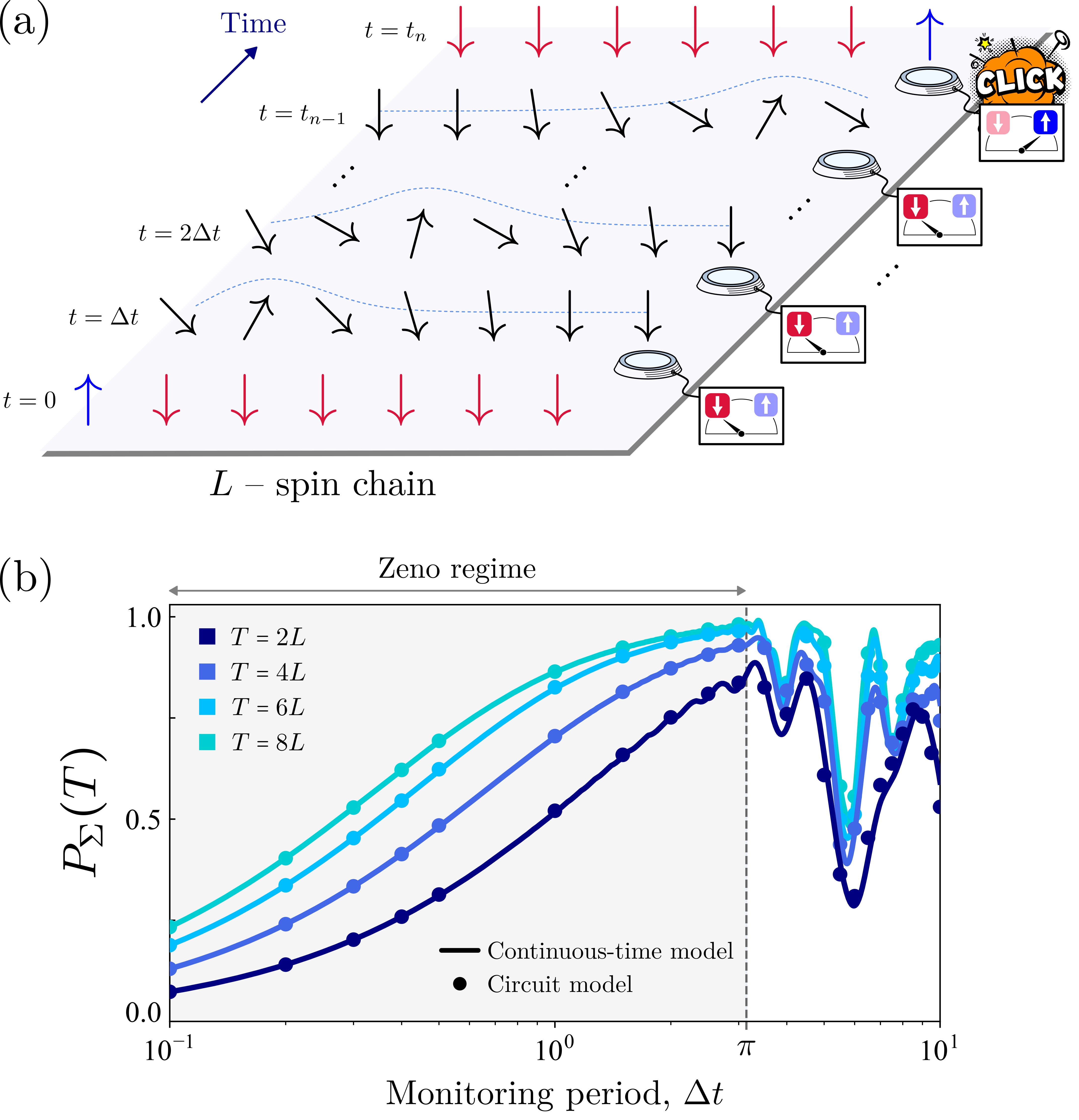}  
        \caption{Arrival-time problem in monitored spin chains and circuits. (a): Single or multiple spin-up sites are prepared on one end of the system. During the evolution, the last site is periodically monitored  until the first spin-up outcome ("detector click"). (b): Cumulative arrival-time probability $P_{\Sigma}(T)$  in the two monitoring regimes. In the quantum Zeno regime, the arrival probability is suppressed due to rapid monitoring, while in the low-monitoring regime the probability becomes non-monotonic. Time is expressed in units of the exchange coupling $J^{-1}$.   }
    \label{fig1}
\end{figure}

Detector clicks could serve as simple practical signatures of arrival times, but predicting their temporal distributions faces several conceptual
challenges. Time, unlike position, is not an operator in standard quantum theory: no self-adjoint time operator is conjugate to a Hamiltonian bounded from below~\cite{Pauli1933}, so candidate arrival-time distributions must be
formulated as generalized measurements or derived from explicit detection models~\cite{Kijowski1974,Allcock1969,MugaLeavens2000}. Furthermore, temporal distributions are sensitive to the details of the detection process, a fact which typically does not factor into the problem of spatial distributions.
Measurements introduce non-unitary dynamics beyond the isolated unitary evolution, and their precise modeling is often case-dependent. The observed arrival-time distribution therefore unavoidably reflects a combination of
unitary propagation and non-unitary monitored dynamics. This can even lead to paradoxical behavior: increasing the temporal resolution of projective monitoring increases the disturbance to the unitary evolution, culminating in
the quantum Zeno regime where frequent monitoring suppresses arrival altogether~\cite{MisraSudarshan1977,Itano1990,Mielnik1994}. Together with these conceptual challenges, the lack of controllable platforms with high spatial and temporal detector resolution has prevented systematic experimental exploration of arrival-time statistics. Digital quantum processors with mid-circuit measurements~\cite{Corcoles2021}, where the
interplay of unitary dynamics and repeated measurements is now studied
extensively~\cite{Li2018,Skinner2019}, offer precisely such a platform.

In the present work, we adopt the repeated-measurement definition of arrival
time on lattices~\cite{Dhar2015JPA,Dhar2015PRA,Friedman2017PRE} and study it
in $U(1)$-symmetric spin chains and quantum circuits under periodic
monitoring. These systems, illustrated in Fig.~\ref{fig1}(a), correspond to particles on a lattice, where an up-spin site represents a particle
and a down-spin site an empty site. The last site of the system is monitored
periodically by projective measurements, and the initial state is prepared so
that the up-spin sites lie at the opposite end from the detector site. As illustrated in Fig.~\ref{fig1}(b), we show
that the arrival-time dynamics exhibits a crossover between two qualitatively
distinct regimes, separated by an optimal monitoring rate: a quantum Zeno
regime, where frequent monitoring suppresses arrival, and a low-monitoring
regime with non-monotonic dependence on the monitoring period. Simple analytic
arguments, a universal Zeno bound and the single-particle bandwidth, account
for both regimes and locate the optimum. We observe this crossover experimentally on superconducting quantum processors using
mid-circuit measurements, obtaining quantitative agreement with quantum
trajectory theory, and extend the measurements to multiple interacting
particles and two-dimensional geometry. To our knowledge, this work provides
the first experimental measurement of arrival-time distributions under repeated projective monitoring in a transport setting.

\section{Arrival time problem in spin chains and quantum circuits}

\subsection{Models}

We study the arrival-time problem in a monitored spin-1/2 XXZ chain and in a
monitored $U(1)$-symmetric quantum circuit, as depicted in Fig.~\ref{fig2}(a). The circuit converges to the
continuous-time chain in the limit of a small time step, but we treat the
time step as an independent parameter, so that the two models offer
complementary advantages. The monitored chain, with its continuous-time
dynamics, provides a direct analogy with the traditionally studied
arrival-time problem, while the monitored circuit realizes the same
phenomenology and is naturally implemented on existing digital quantum
hardware.

The Hamiltonian of the XXZ chain is
\begin{equation}
H=\frac{J}{4}\sum_{i=1}^{L-1}\left(X_iX_{i+1}+Y_iY_{i+1}+\Delta Z_iZ_{i+1}\right),
\label{eq:H}
\end{equation}
where $J$ denotes the exchange coupling, $\Delta$ is the anisotropy parameter,
and $X_i$, $Y_i$ and $Z_i$ are the Pauli operators acting on site $i$ of a chain
of $L$ sites. In the following, we express time in units of $J^{-1}$ and set $J = 1$.
The total magnetization is conserved, $[\sum_i Z_i,H]=0$, which makes spin
transport through the chain a well-defined concept. We consider initial states
\begin{equation}\label{eq:initialstate}
|\psi(t=0)\rangle=|\uparrow\cdots\uparrow\downarrow\downarrow\cdots\downarrow\rangle,
\end{equation}
where $k$ up-spins are prepared at one end of the chain and the remaining
$L-k$ sites are down. Identifying up-spins with particles, the model
describes hard-core bosons with nearest-neighbor hopping $J/2$ and
nearest-neighbor interaction $J\Delta$ (up to a boundary potential); in one
dimension, a Jordan--Wigner transformation~\cite{JordanWigner1928} maps them
to interacting spinless fermions (see App.~\ref{app:optimal} for details). The last site is monitored by projective
measurements with period $\Delta t$, implemented by the projectors
$\hat P_\uparrow=|\uparrow\rangle\langle\uparrow|$ and
$\hat P_\downarrow=|\downarrow\rangle\langle\downarrow|$ onto the eigenstates
of $Z_L$. The time evolution thus consists of unitary evolution
$U(t)=e^{-iHt}$, periodically interrupted by measurements, as in the
stroboscopic first-detection protocol~\cite{Dhar2015PRA,Friedman2017PRE}.
Denoting $t_n=n\Delta t$, the state immediately after the $n$th measurement,
with outcome $\sigma_n\in\{\uparrow,\downarrow\}$, is
\begin{equation}
|\tilde\psi_{\bm m_n}(t_n)\rangle=\hat P_{\sigma_n}U(\Delta t)\,
|\psi_{\bm m_{n-1}}(t_{n-1})\rangle ,
\label{eq:traj}
\end{equation}
normalized as $|\psi_{\bm m_n}\rangle=|\tilde\psi_{\bm m_n}\rangle/
\||\tilde\psi_{\bm m_n}\rangle\|$. The monitored dynamics defines a quantum
trajectory $|\psi_{\bm m_n}(t)\rangle$~\cite{Dalibard1992,WisemanMilburn2009},
where the record $\bm m_n=(\sigma_1,\dots,\sigma_n)$ collects the outcomes of
the first $n$ measurements, which occur with Born-rule probabilities. Below,
trajectories with specific records are employed to construct the arrival-time
distributions.

In the quantum circuit model, the monitored dynamics follows the same
evolution~\eqref{eq:traj}, with $U(\Delta t)$ replaced by $U_\delta^{\,m}$,
where $\Delta t=m\,\delta t$ and
\begin{equation}
U_\delta=\prod_{i\,\mathrm{even}}e^{-ih_{i,i+1}\delta t}
\prod_{i\,\mathrm{odd}}e^{-ih_{i,i+1}\delta t}
\label{eq:circuit}
\end{equation}
is one brickwork cycle composed of the two-qubit gates
\begin{equation}
h_{i,i+1}=\frac{J}{4}\left(X_iX_{i+1}+Y_iY_{i+1}+\Delta Z_iZ_{i+1}\right),
\label{eq:h}
\end{equation}
so that $H=\sum_{i}h_{i,i+1}$. Gates within each product act on disjoint
pairs and commute. Here $\delta t$ is the effective time step, and the time
after $N$ cycles is $t=N\delta t$. The circuit, depicted in
Fig.~\ref{fig2}(a), is the first-order Trotterization of
$e^{-iH\delta t}$~\cite{Lloyd1996}, which approximates it accurately for
$J\delta t\ll1$. For finite $\delta t$, the brickwork circuit is itself an
integrable Floquet model with ballistic spin
transport~\cite{Vanicat2018,Ljubotina2019}, and can be regarded as a
discrete-time counterpart of the spin chain in its own right. Both models
propagate a conserved $U(1)$ charge, initially localized at one end, towards
the monitored site at the other end, and the arrival-time problem under
periodic monitoring is formulated identically for both.

\subsection{The arrival-time problem} \label{subsec:arrivaltime}

\begin{figure*}
    \centering
    \includegraphics[width=1.9\columnwidth]{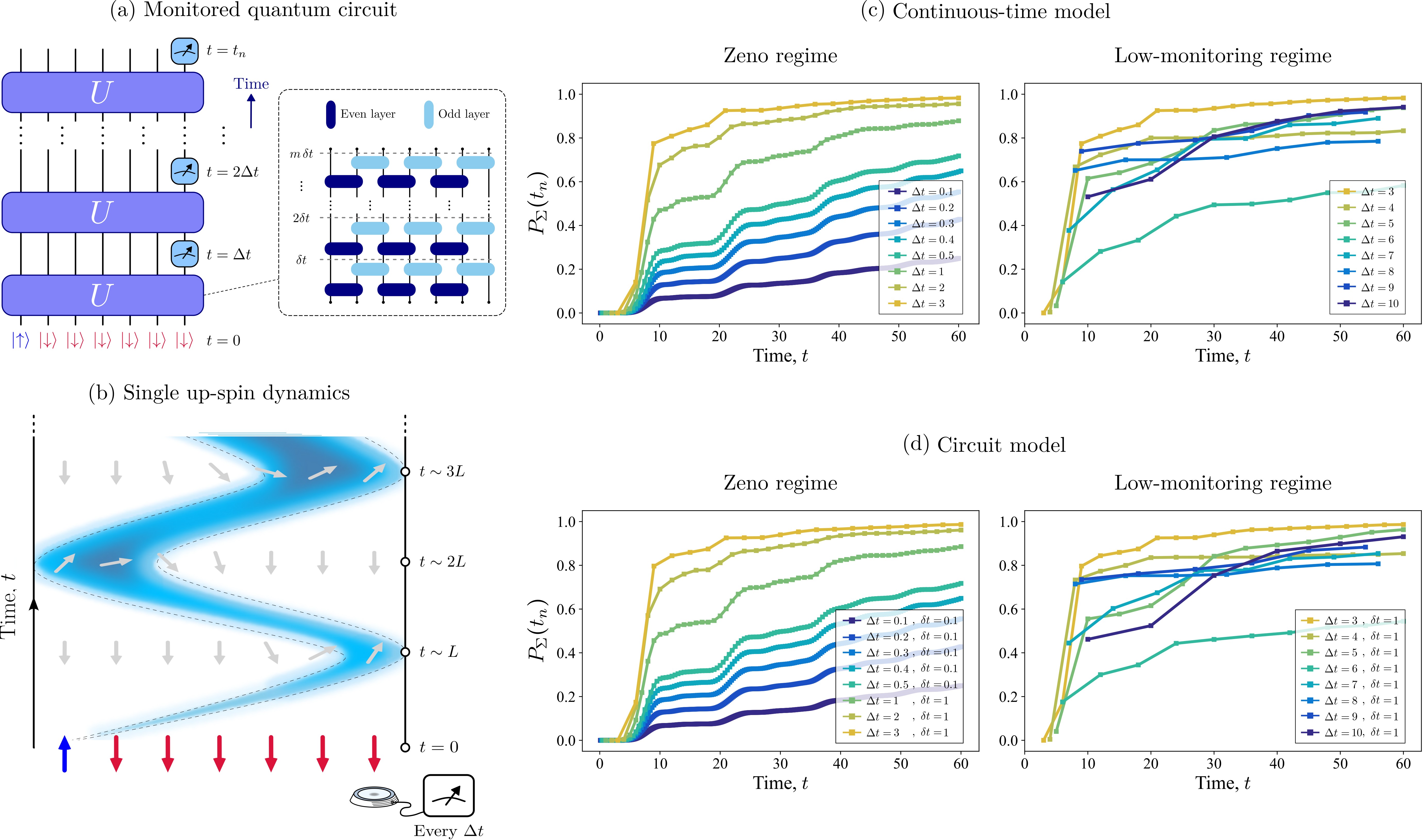}  
         \caption{Arrival-time problem in different monitoring regimes for $L = 7$. (a):  Schematic presentation of the monitored dynamics. Temporal evolution between monitoring events can be generated by a Hamiltonian $U=e^{-iHt}$ or layers of $U(1)$-symmetric two-qubit gates. (b): Schematic evolution of a single spin-up initial state. Arrival probabilities are enhanced around times that are odd multiples of $L$. (c): Cumulative arrival probability for a single up-spin site in the quantum Zeno and low-monitoring regimes of the continuous-time model with anisotropy $\Delta=1/2$. (d): Same as (c) but for the circuit model with the Trotter steps
        $\delta t$.}
    \label{fig2}
\end{figure*}

 The quantity of interest is the probability of the first observation of an up-spin particle at the $n$th monitoring event at $t_n=n\Delta t$. This can be constructed from quantum trajectories with specific measurement records. We denote the probability of observing an up-spin particle at the monitoring site at $t_n$, when all the previous measurements have resulted in down-spin observations, by $p_\uparrow(t_n)$. This is given by
\begin{equation*}
 p_\uparrow(t_n)=\langle \psi_{{\bm m}_{n-1}}(t_{n-1})|U^\dagger(\Delta t)\hat{P}_{\uparrow} U(\Delta t) | \psi_{{\bm m}_{n-1} }(t_{n-1})\rangle,   
\end{equation*}
where the trajectory which yields the correct probability corresponds to the measurement record of $n-1$ down-spin measurements ${\bm m}_{n-1}=\{\downarrow,\downarrow\ldots \downarrow\}$. Similarly, the probability of measuring a down-spin at time $t_n$, when all the previous measurements have already resulted in spin-down observations, is given by 
$p_\downarrow(t_n)=1-p_\uparrow(t_n)$. The arrival probability $P_a(t_n)$, defined as the first observation of up-spin particle at $t_n$, is obtained as
\begin{align}\label{eq:arrivalprob}
 P_a(t_n)=\left[\prod_{n'<n} p_\downarrow(t_{n'})\right] [1-p_\downarrow(t_n)]
\end{align}
which can be evaluated by solving the quantum trajectory with the all-down measurement record. The result \eqref{eq:arrivalprob} has a simple interpretation; factor $[1-p_\downarrow(t_n)]$ corresponds to the probability of observing up-spin in the $n$th measurement, \emph{provided} that all the previous measurements observed spin down, while the factor $\prod_{n'<n} p_\downarrow(t_{n'})$ yields precisely the probability that all the earlier measurements resulted in spin-down observations. The cumulative arrival-time probability, defined as the probability that the first observation of up-spin at the monitoring site happens by the time $t_n$, is given by 
\begin{align}\label{eq:cumlativeprob}
 P_\Sigma(t_n)=\sum_{i=1}^n P_a(t_i).
\end{align}
The survival probability, defined as $P_s(t_n)=1-P_\Sigma(t_n)$, yields the probability that no up-spin particle has been detected by the time $t_n$. The arrival probability and the survival probability are connected by the recursive relation $P_a(t_n)=P_s(t_{n-1})-P_s(t_{n})$, which yields an asymptotic relation $P_\Sigma(\infty)=1-P_s(\infty)$. In a finite geometry, in the absence of localization and of dark states at exceptional monitoring periods (App.~\ref{app:optimal}), one expects that $P_s(\infty)=0$, so that the particle eventually arrives at the monitoring site with unit probability.       

In Fig.~\ref{fig2} we illustrate the cumulative arrival-time distributions starting from the initial state \eqref{eq:initialstate} with a single spin-up site. Fig.~\ref{fig2}(b) introduces the characteristic time scales relevant for the problem. The arrival probability becomes finite after the up-spin particle has had time to traverse the system which happens around $t\sim L$. If the particle goes undetected then, it is reflected back and a period of low-arrival probability follows for the duration $L \lesssim t\lesssim 3L$, corresponding to a round trip of the wave packet. At longer times, a similar pattern repeats but attenuated, as the wave packet broadens during the evolution. These features are clearly visible in the high-frequency monitoring regime in the continuous-time model, as seen in Fig.~\ref{fig2}(c) in the left panel. Reflecting the quantum Zeno effect, the arrival-time probability at high monitoring frequency is strongly suppressed and grows monotonically as the monitoring period increases, until reaching maximum detection efficiency close to $\Delta t\sim 3$. As shown in App.~\ref{app:optimal}, while the optimum point is close to $\Delta t= \pi$, it is achieved in a very flat minimum, defining a broad (nearly) optimal region. As the Fig.~\ref{fig2}(c) right panel shows, increasing the monitoring period beyond the optimal region, the system enters the low-monitoring regime, characterized by strikingly non-monotonic dependence of arrival probability on the monitoring period. The non-monotonicity arises from commensurability effects between the periodic wave packet motion and the periodic monitoring. Thus, the optimal monitoring period divides the arrival-time dynamics into two qualitatively distinct dynamical regimes, the quantum Zeno regime and the low-monitoring regime. Both regimes and their boundary follow from simple analytic arguments (App.~\ref{app:optimal}). For any state, anisotropy and particle number, the click probability per measurement obeys $p_\uparrow\le J^2\Delta t^2/4$, so arrival is suppressed as $\Delta t\to0$. For long periods, stroboscopic resonances $(E_a-E_b)\Delta t=2\pi\ell$ between single-particle levels~\cite{Friedman2017PRE,Thiel2020} create nearly dark, slowly decaying modes, which first become possible at $\Delta t\approx2\pi/W=\pi/J$, with $W=2J$ the bandwidth, in agreement with the observed optimum. As seen in Fig.~\ref{fig2}(d), the circuit model exhibits very similar arrival-time distributions to the continuous-time model, including the same optimum. It should be noted that, while the circuit model requires a sufficiently short effective step $\delta t\ll 1$ in order to probe the deep Zeno regime, the crossover at the optimal monitoring regime is observable even for a large step $\delta t=1$. In App.~\ref{app:optimal}, the above results are also illustrated for multiple spin-up particles, which map to interacting multi-particle systems. While the detailed form of the arrival-time distributions is significantly modified, the two distinct measurement-induced dynamical regimes are qualitatively preserved even in the presence of many-body effects.

\section{Observing arrival-time distributions on quantum hardware}
\label{sec:experiment}

\begin{figure*}
    \centering
    \includegraphics[width=2\columnwidth]{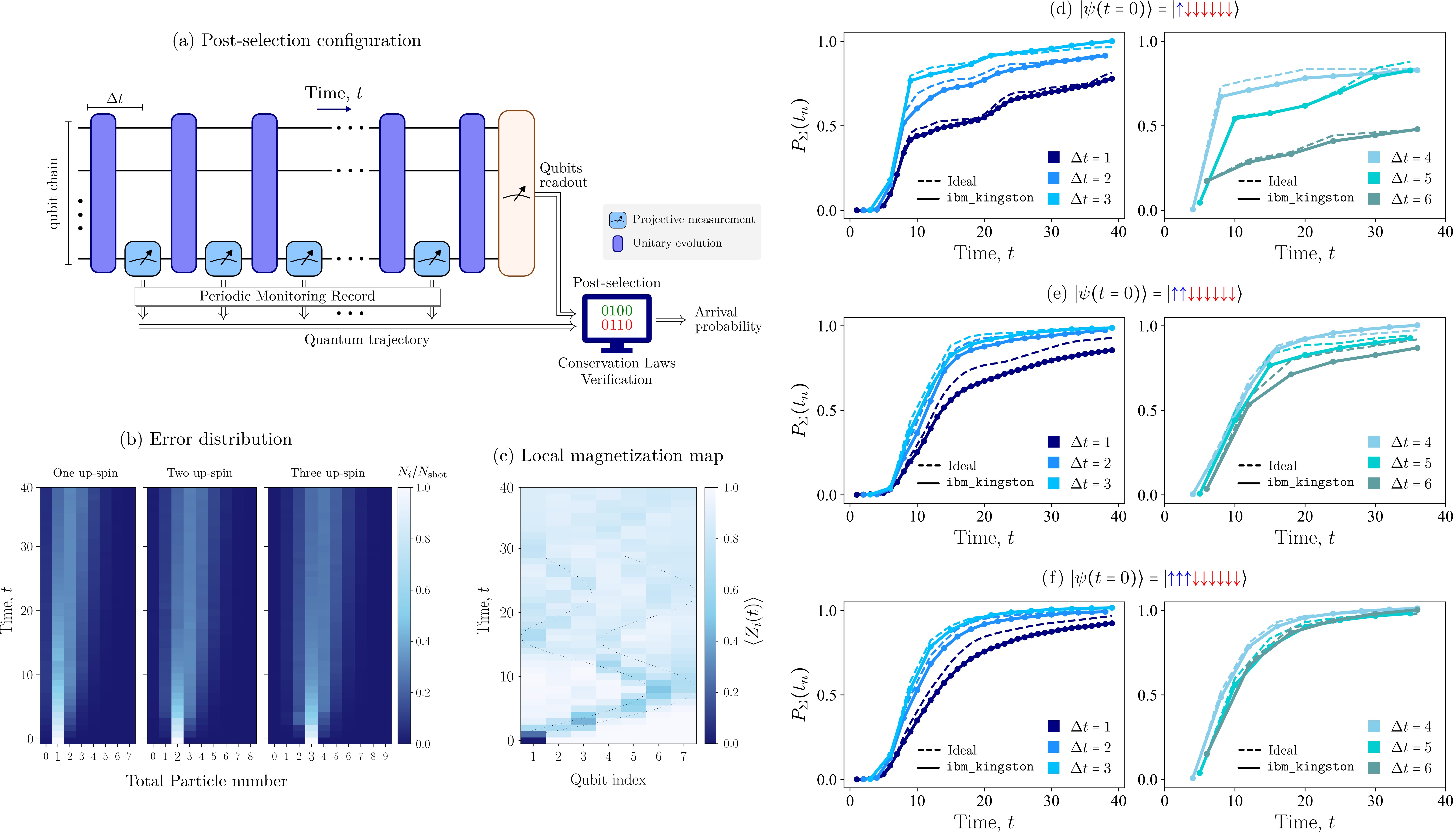}  
        \caption{ Measuring arrival-time distributions on a digital quantum computer. (a): Setup and the error-mitigation scheme. (b): Symmetry-breaking noise drives the state from the initial $U(1)$-sector during the evolution. The map shows the fraction of shots found in each total particle-number sector at time $t$. The weight remaining in the initial sector equals the post-selected fraction $N_i/N_{\rm shot}$, so the error rate is $1-N_i/N_{\rm shot}$.
         The noise-induced particle non-conservation is mitigated by post-selection into the correct $U(1)$ sector. (c): Local magnetization map at different times for a single up-spin initial state and $\Delta t=1$. (d): Arrival-time distributions around the quantum Zeno crossover for a single up-spin particle.  (e) and (f) are the same as (d) but for two and three up-spin particles in the initial state. The data have been obtained for the anisotropy parameter $\Delta=1/2$, corresponding to interacting up-spin excitations.}
    \label{fig3}
\end{figure*}

While the arrival-time dynamics of the continuous-time and circuit models
exhibit qualitatively the same features, the latter is better suited for
current digital quantum computers. Since the evolution time is $t=N\delta t$
and an accurate simulation of the continuous-time model requires
$\delta t\ll1$, reaching long times translates into large circuit depth, at
which noise unavoidably accumulates on current processors. However, as noted
above, the crossover between the quantum Zeno and low-monitoring regimes
remains accessible in the circuit model for a relatively large time step
$\delta t=1$. This is the sweet spot where the arrival-time dynamics can be
probed around the crossover at depths that still allow efficient error
mitigation.

\subsection{Arrival-time distributions at the quantum Zeno crossover}
\label{sec:1d}

We implement the circuit model of Fig.~\ref{fig2}(a) with the brickwork
cycles~\eqref{eq:circuit} on IBM's superconducting quantum processor
\texttt{ibm\_kingston}~\cite{IBMQuantum} (see Appendix~\ref{app:hardware}).
After preparing the initial state~\eqref{eq:initialstate}, we carry out the unitary
evolution interrupted by mid-circuit measurements~\cite{Corcoles2021}. By
repeating the monitored evolution sufficiently many times, we directly obtain
the arrival probabilities $P_a(t_n)$ and the cumulative probability
$P_\Sigma(t_n)$. While simulating the model is in principle straightforward,
a naive execution is soon overwhelmed by noise on present-day hardware, and
quantitatively accurate results require noise suppression and error
mitigation~\cite{Nation2021M3,ViolaKnillLloyd1999,BonetMonroig2018}. The two
leading error sources are $U(1)$-symmetry-breaking noise, arising from gate
errors and qubit relaxation, and decoherence during the long mid-circuit
measurements. Compared to the typical two-qubit gate duration of 68~ns, a
mid-circuit measurement takes 2180~ns, during which the idle qubits
decohere significantly. This is strongly suppressed by dynamical
decoupling~\cite{ViolaLloyd1998,ViolaKnillLloyd1999} of the idle qubits
during the measurement. The $U(1)$-breaking noise requires a different
approach, as it changes the total $U(1)$ charge during the evolution; if
significant, the monitored statistics no longer faithfully reflect the
arrival statistics of particles that have traversed the system. Moderate
$U(1)$ breaking can be mitigated by symmetry-based
post-selection~\cite{BonetMonroig2018,McArdle2019}, illustrated in
Fig.~\ref{fig3}(a). The circuit is executed separately up to each
monitoring time $t_n$, after which all qubits are measured. The same runs
yield both the first up-spin detection at the last site and the total spin
at the termination depth, and runs whose total spin differs from the
initially prepared one are discarded. As shown in Fig.~\ref{fig3}(b), the
probability of ending up in an incorrect sector grows rapidly with circuit
depth, so the post-selection (Appendix~\ref{app:postselection}) adds a large
sampling overhead at late times. Noise processes that preserve the total
spin, such as compensating spin flips accumulated over many gate layers, are
not removed by this procedure and constitute the dominant residual error at
late times.

In Fig.~\ref{fig3}(c) we plot the local magnetization as a function of the
final measurement time for the monitored circuit. The up-spin wave packet
first propagates ballistically towards the monitoring site; after being
reflected, it broadens strongly and becomes highly delocalized after one
round trip. The cumulative arrival-time distributions for a single up-spin,
shown in Fig.~\ref{fig3}(d), agree closely with the theoretical prediction. The absolute statistical uncertainty of the measured curves, characterized by the standard deviation, is around $1\%$, which is much smaller than the characteristic features
(see Appendix~\ref{app:statistics} for details). The left panel in Fig.~\ref{fig3}(d) illustrates the quantum Zeno regime, where the arrival probability is monotonically increasing as the monitoring period is increased $\Delta t=1\to3$ towards the optimal point.  When the monitoring period is increased beyond the optimal region $\Delta t=3\to6$, as seen in the right panel, the arrival probability decreases and the system moves into the low-monitoring regime. Remarkably, the experiment reproduces accurately the
periodic fine structure arising from the wave-packet round trips. These experimental observations confirm the key predictions of the quantum-trajectory theory, most importantly the existence of the distinct dynamical regimes with their distinct characteristic features.

In Figs.~\ref{fig3}(e) and (f) we show how adding multiple up-spin sites, corresponding to multiple interacting particles, affects the arrival-time distributions. The system size is increased, so that the
initial distance between the closest up-spin and the detector site is kept fixed. This makes the arrival time behavior conveniently comparable to the single particle case. As predicted by the theory, $\Delta t=3$ again lies very close to the optimal monitoring point, exhibiting the highest detection probability among the monitoring periods studied. At late times
$t\gg L$, the first arrival probability for two and three spin-up states is substantially enhanced compared to the single-spin case, as any of the particles can trigger the detector. Furthermore, in agreement
with theory (see App.~\ref{app:optimal}), the plateaus arising from wave-packet round trips are suppressed, since multiple interfering many-body excitations no longer produce periods of sharp suppression at the detector. For multiple spin-up states, the deviations from the
theory are larger than in the single-particle case. As Fig.~\ref{fig3}(b)
shows, $U(1)$-breaking errors grow with the number of up-spins. Moreover,
the number of noise processes that return the system to the initial sector
through compensating spin flips grows towards half filling, so a larger
fraction of corrupted runs survives the post-selection.

\subsection{Arrival-time distributions in 2d geometry}

\begin{figure*}
    \centering
    \includegraphics[width=2
    \columnwidth]{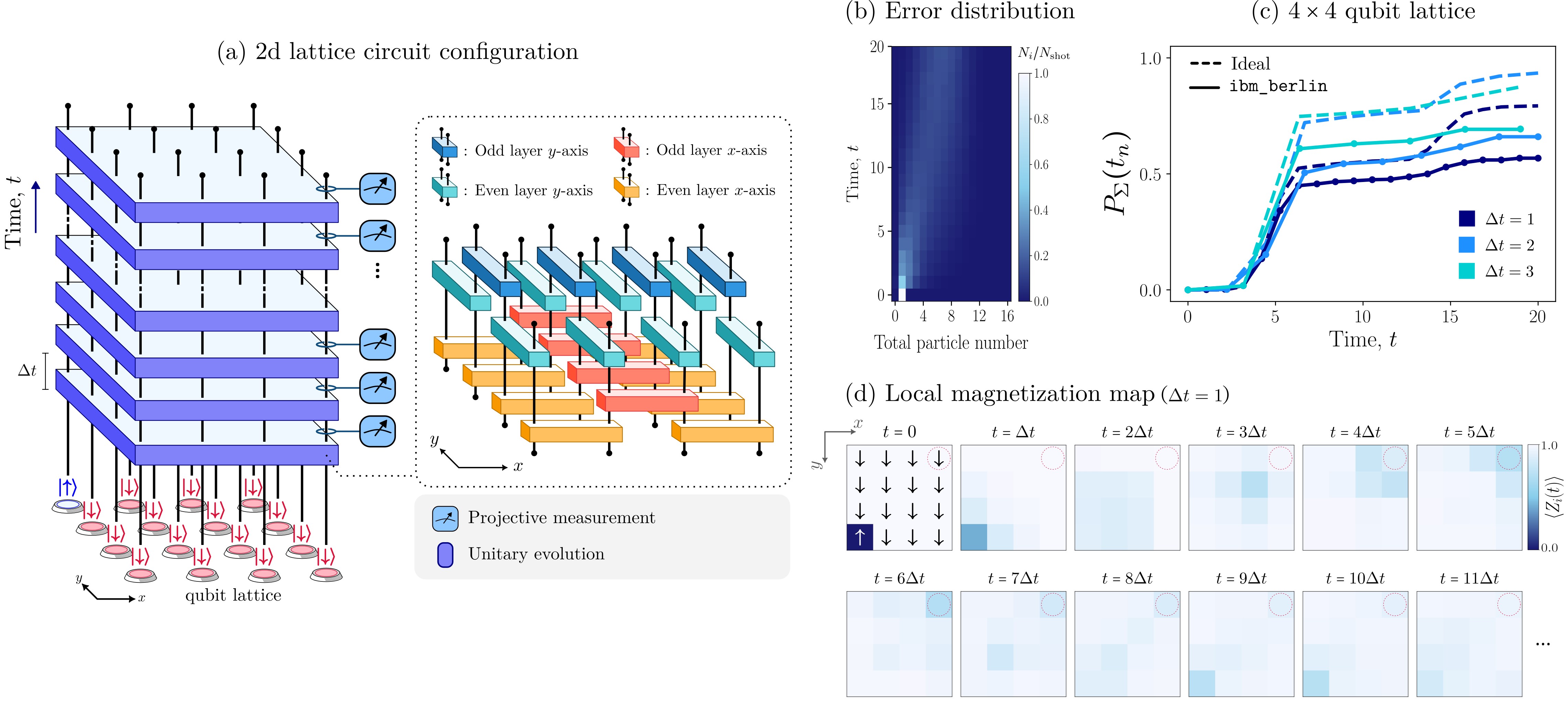}  
        \caption{Arrival-time distributions in 2d geometry. (a): Circuit consists of a $4\times 4$ square lattice, where the starting location and monitored site reside in opposite corners. The 2d unitary employs the same building blocks as the 1d circuit. (b): Evolution of the $U(1)$ sector due to the symmetry-breaking noise in the circuit. (c): Obtained arrival-time distributions after error mitigation for different monitoring periods. (d): Evolution of the local magnetization. }
    \label{fig4}
\end{figure*}

We next test whether the two regimes survive in higher dimensions. The 1d circuit model generalizes to a 2d square lattice of
$L\times L$ sites labeled by $\mathbf{r}=(x,y)$. As illustrated in Fig.~\ref{fig4}(a), each unitary cycle consists of the 1d brickwork circuit \eqref{eq:circuit} along rows and columns,
\begin{equation}
U_{{\rm 2d},\delta}=U_y^{\rm even}\,U_y^{\rm odd}\,
U_x^{\rm even}\,U_x^{\rm odd},
\label{eq:U2d} \nonumber
\end{equation}
where
\begin{equation}
U_x^{\rm odd/even}=\prod_{y=1}^{L}\;\prod_{x\ \rm odd/even}
e^{-i h_{(x,y),(x+1,y)}\delta t},\nonumber
\end{equation}
\begin{equation}
U_y^{\rm odd/even}=\prod_{x=1}^{L}\;\prod_{y\ \rm odd/even}
e^{-i h_{(x,y),(x,y+1)}\delta t}. \nonumber
\end{equation}
Here $h_{\mathbf{r},\mathbf{r}'}$ is the two-qubit operator \eqref{eq:h}, now acting on nearest-neighbor pairs along the $x$ and
$y$ directions. Since every gate conserves the total magnetization, the arrival-time problem is defined exactly as before. We implement the 2d circuit on \texttt{ibm\_berlin}, whose square-lattice connectivity avoids SWAP overhead (App.~\ref{app:hardware}), prepare a single spin-up site in one corner $\mathbf{r}_0=(1,1)$, and monitor the opposite corner
$\mathbf{r}_{\rm m}=(L,L)$. The evolution of the $U(1)$ charge obtained in the error-mitigation procedure is shown in Fig.~\ref{fig4}(b), illustrating that the significantly increased number of gate operations in the 2d model gives rise to strongly enhanced noise effects compared to 1d model. Thus, the correspondence between the observed and ideal arrival-time distributions, presented in Fig.~\ref{fig4}(c), becomes less precise compared to the 1d case. Nevertheless, the Zeno regime persists in 2d: the highest monitoring frequency suppresses the arrival probability, as predicted by the dimension-independent bound $p_\uparrow\le J^2\Delta t^2/4$. The bouncing wave-packet motion, which gives rise to the plateaus in the arrival-time distributions, is clearly seen in the magnetization map in Fig.~\ref{fig4}(d). While the overall phenomenology is similar to that of the 1d model, the quicker accumulation of noise in the 2d system restricts the quantitative match with theory to a narrower time window.

\section{ Discussion and outlook}

Despite nearly a century of theoretical work on the quantum arrival-time
problem~\cite{Allcock1969,Kijowski1974,MugaLeavens2000}, experiments resolving
arrival-time distributions of particles propagating through a system to a
detector have remained lacking, largely because no platform combined
controllable spatial propagation with high-resolution, repeated monitoring of a
fixed detector site. Related experiments have measured quantum first-passage
statistics in non-spatial settings, such as the energy of a trapped
ion~\cite{Ryan2025} and the transferred charge in mesoscopic
conductors~\cite{Singh2019}, while the quantum Zeno effect has been observed
in contexts unrelated to particle arrival~\cite{Itano1990,Schafer2014}.

On the theory side, the stroboscopic first-detection protocol for
tight-binding lattices was developed by Friedman, Kessler and
Barkai~\cite{Friedman2017PRE,Friedman2017JPA}. They predicted Zeno suppression
at short sampling times, as well as divergences and resonances at exceptional
sampling times associated with wave-packet revivals. First-detection times of
monitored quantum walks have since been measured on superconducting processors
with mid-circuit readout, for minimal two- and three-site
graphs~\cite{TornowZiegler2023,Wang2024}, focusing on mean return times, their
topological quantization and dark states. Measurement-induced effects on the
recurrence of photonic quantum walks have also been
demonstrated~\cite{Nitsche2018}. Our work differs from these studies in four
respects. First, it addresses arrival in a transport geometry, where a
particle prepared at one end of an extended system propagates to a monitored
site at the other end. Second, it resolves full arrival-time distributions
rather than mean detection times. Third, it shows that in this geometry the
Zeno suppression and the revival-related resonances organize into two
qualitatively distinct regimes separated by an optimal monitoring rate: a Zeno
regime, where detection increases monotonically with the monitoring period,
and a low-monitoring regime, where commensurability between the round-trip
motion of the wave packet and the monitoring period produces strongly
non-monotonic arrival statistics. Fourth, it shows that this organization
survives beyond the single-particle chain: with two and three interacting
particles the optimal monitoring period is preserved while many-body
interference reshapes the distributions, and in a two-dimensional lattice the
Zeno suppression persists.

Our results also bear on the long-standing question of which arrival-time
distribution quantum mechanics predicts. Mielnik's ``cannonball'' argument
invokes the Zeno effect to question whether repeated projective measurements
can yield arrival times at all~\cite{Mielnik1994}; the Zeno regime observed
here realizes this effect directly in a transport setting. According to a modern view, alternative arrival-time distributions are not competing predictions for
a single observable but correspond to operationally distinct
protocols~\cite{Roncallo2023,Martellini2025}. In the recently proposed quantum
stroboscopy~\cite{Lloyd2025}, each copy of the system is measured only once,
at a different time, which bypasses the Zeno effect and reconstructs an
undisturbed arrival-time distribution. In contrast, our protocol monitors the
same system repeatedly, treating the measurement back-action as an intrinsic
part of the arrival process, and it is this back-action that gives rise to the
two dynamical regimes. Since the stroboscopic protocol requires only omitting
the mid-circuit measurements and reading out the detector site at the final
time, both distributions, along with other operational definitions
formulated for spin chains~\cite{Martellini2025}, could be studied on the same
hardware. In App.~\ref{app:unmonitored} we present a theoretical comparison between the unmonitored arrival-time probabilities and our prescription. This comparison emphasizes how arrival-time distributions depend on the measurement procedure and the
conditioning used to define them.

\section{Summary}

We report, to our knowledge, the first experimental measurement of arrival-time distributions in a monitored quantum transport setting, realized on superconducting quantum processors with mid-circuit measurements. In periodically monitored $U(1)$-symmetric quantum circuits, we observe a crossover between two theoretically predicted dynamical regimes: a quantum Zeno regime where frequent measurement suppresses arrival, and a low-monitoring regime with striking non-monotonic dynamics, separated by an optimal monitoring period set by the single-particle bandwidth. This crossover is not a single-particle effect. For two and three interacting particles, the optimal monitoring period is preserved, while many-body interference washes out the round-trip plateaus and enhances late-time detection. In a two-dimensional lattice, frequent monitoring continues to suppress arrival. This indicates that the two regimes are a generic consequence of repeatedly monitoring a propagating system, rather than a feature of a single particle in a chain. Our work brings one of the longest-standing open problems in quantum mechanics onto a controlled, scalable platform. It establishes monitored quantum circuits as a powerful tool for investigating fundamental questions in quantum dynamics.

\subsection*{Data Availability}
The source code, simulation parameters and experimental results generated and/or analyzed during the current study are available in the GitHub repository \cite{MyCode2026}.

\acknowledgements
This work was supported by the Finnish Research Council project 362573 and the Finnish quantum flagship program. K.P. acknowledges the support by the Finnish Research Council project 363879. This work is part of the Finnish Center of Excellence in Quantum Materials (QMAT).

 We acknowledge the use of IBM Quantum Credits for this work. The views expressed are those of the authors, and do not reflect the official  policy or position of IBM or the IBM Quantum team.

\appendix

\section{Optimal monitoring and many-body effects }\label{app:optimal}

At $\Delta=0$ the monitored chain maps onto free fermions, and the arrival-time
problem can be solved exactly for 1d chain with any number $k$ of up-spin particles  and any system size $L$. Below we first derive the average first detection time and then use it to
determine the optimal monitoring period for that quantity. While the first detection time is not directly related to the cumulative arrival-time probability we study, for the systems considered the optimal monitoring periods for both quantities nearly coincide, as discussed below.

\subsection{First detection time}

Under the Jordan--Wigner transformation an up-spin on site $j$ becomes a
fermion, $n_j=c_j^\dagger c_j$. At $\Delta=0$ the Hamiltonian is quadratic,
$H=\sum_{ij}h_{ij}\,c_i^\dagger c_j$, with $h$ the $L\times L$ tight-binding
matrix of hopping amplitude $J/2$. The initial state~(\ref{eq:initialstate})
is a Slater determinant $|\Phi(A_0)\rangle$, whose $k$ occupied orbitals form
the columns of the $L\times k$ matrix $A_0=(e_1,\dots,e_k)$.

Two simple facts make the monitored dynamics tractable. First, the unitary
evolution maps a Slater determinant with orbitals $A$ onto one with orbitals
$u(t)A$, where $u(t)=e^{-iht}$ is an $L\times L$ matrix. Second, the no-click
projector $Q=\hat P_\downarrow$ on the detector site removes exactly the
configurations that occupy site $L$, i.e., it maps $A$ onto $\Pi A$ with
$\Pi=1-|L\rangle\langle L|$. The unnormalized trajectory with record
$\{\downarrow,\dots,\downarrow\}$ therefore remains a Slater determinant,
\begin{equation}
  |\tilde\psi_n\rangle=\big[Q\,U(\Delta t)\big]^n|\psi(0)\rangle
  =|\Phi(A_n)\rangle,\qquad A_n=\mathcal M^{\,n}A_0,
\label{eq:D1}
\end{equation}
where $\mathcal M=\Pi\,u(\Delta t)$ is the single-particle monitored evolution.
The squared norm of $|\tilde\psi_n\rangle$ is the probability that all the measurements up to $t_n=n\Delta t$ have produced no click. Since the norm of a Slater determinant is the
Gram determinant of its orbitals, we obtain
\begin{equation}\label{eq:Ps}
  P_s(t_n)=\langle\tilde\psi_n|\tilde\psi_n\rangle=\det\!\big[A_n^\dagger A_n\big].
\end{equation}
The whole many-body problem thus reduces to products of $L\times L$ matrices.

The arrival probability follows from the survival probability as
$P_a(t_n)=P_s(t_{n-1})-P_s(t_n)$. Summing by parts, the mean arrival time
becomes
\begin{equation}\label{eq:arrival-time}
  \langle t\rangle=\sum_{n\ge1}t_n P_a(t_n)
  =\Delta t\sum_{n\ge0}P_s(t_n)
  =\Delta t\sum_{n\ge0}\det\!\big[A_n^\dagger A_n\big].
\end{equation}

\begin{figure}
    \centering
    \includegraphics[width=0.8\linewidth]{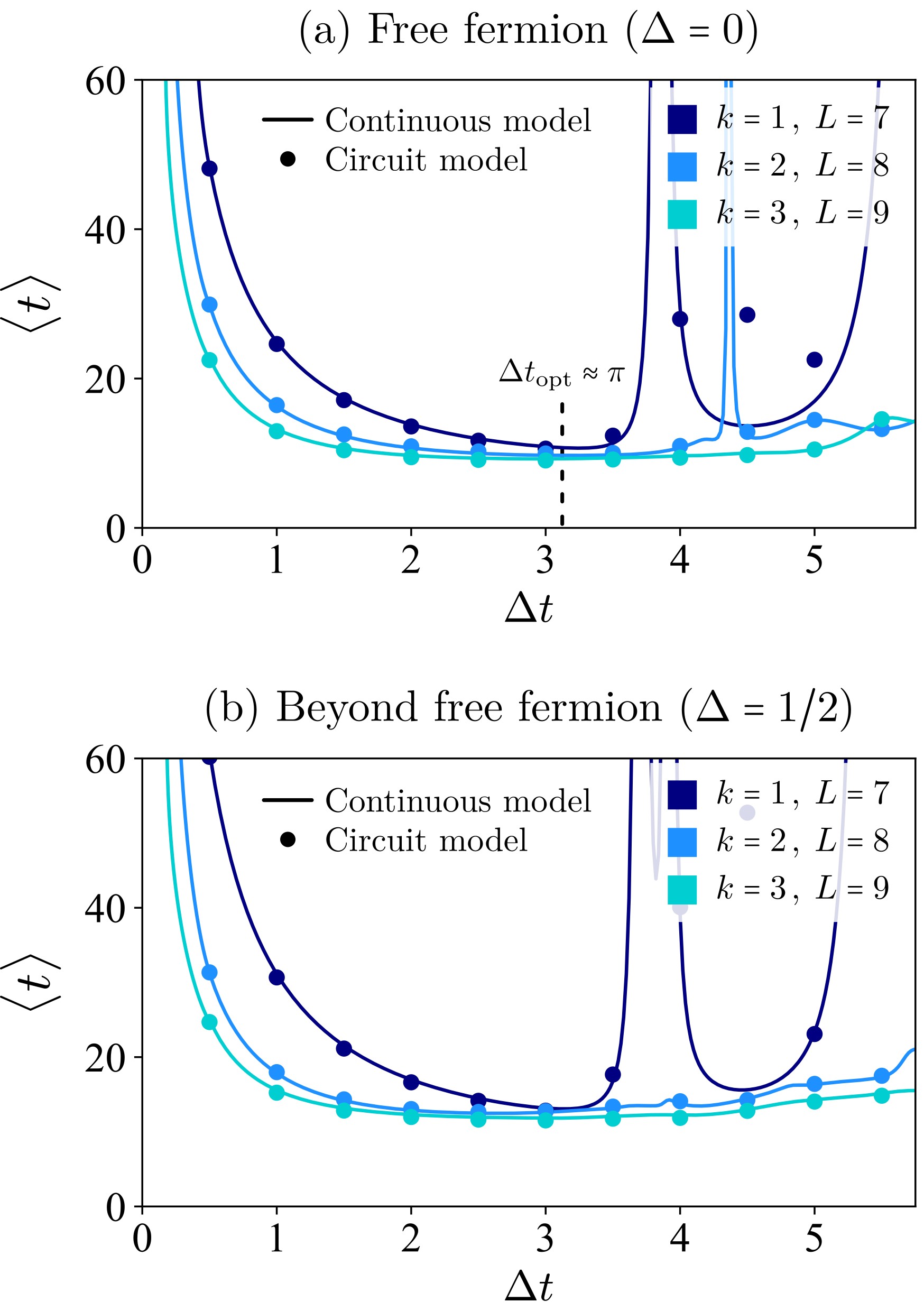}
    \caption{Mean first detection time and the optimal monitoring period.
(a): Mean first detection time $\langle t\rangle$, Eq.~\eqref{eq:arrival-time}, as a
function of the monitoring period $\Delta t$ in the free-fermion limit
$\Delta=0$, for $k=1,2,3$ up-spins prepared at one end of chains with
$L=7,8,9$ sites. The closest up-spin is always six sites from the monitored site. Solid lines show the continuous-time model and dots the
circuit model with large discretization step $\delta t=0.5$. The dashed line marks the estimate
$\Delta t_{\rm opt}\approx\pi$.
(b): Same as (a) but for interacting particles with anisotropy $\Delta=1/2$.}
    \label{fig:6}
\end{figure}

Equation~\eqref{eq:arrival-time} turns the arrival-time problem into a
single sum over determinants, which makes the dependence on the monitoring
period directly accessible (see Fig.~\ref{fig:6}(a)). Two features stand out.
First, the optimal period is essentially independent of the number of
up-spins: adding particles slightly shortens the first detection time, since each one
contributes its own detection channel, but the minimum stays pinned near
$\Delta t\approx\pi$. This is the first indication, confirmed below more generally, that the optimum is effectively determined by
single-particle properties rather than many-body effects.
Secondly, the Trotter circuit reproduces the continuous-time behavior accurately even
for a large time step $\delta t=1$, with its optimum at the closest allowed
period, $\Delta t=3$. Both observations are explained shortly.

\begin{figure*}
    \centering
    \includegraphics[width=1\linewidth]{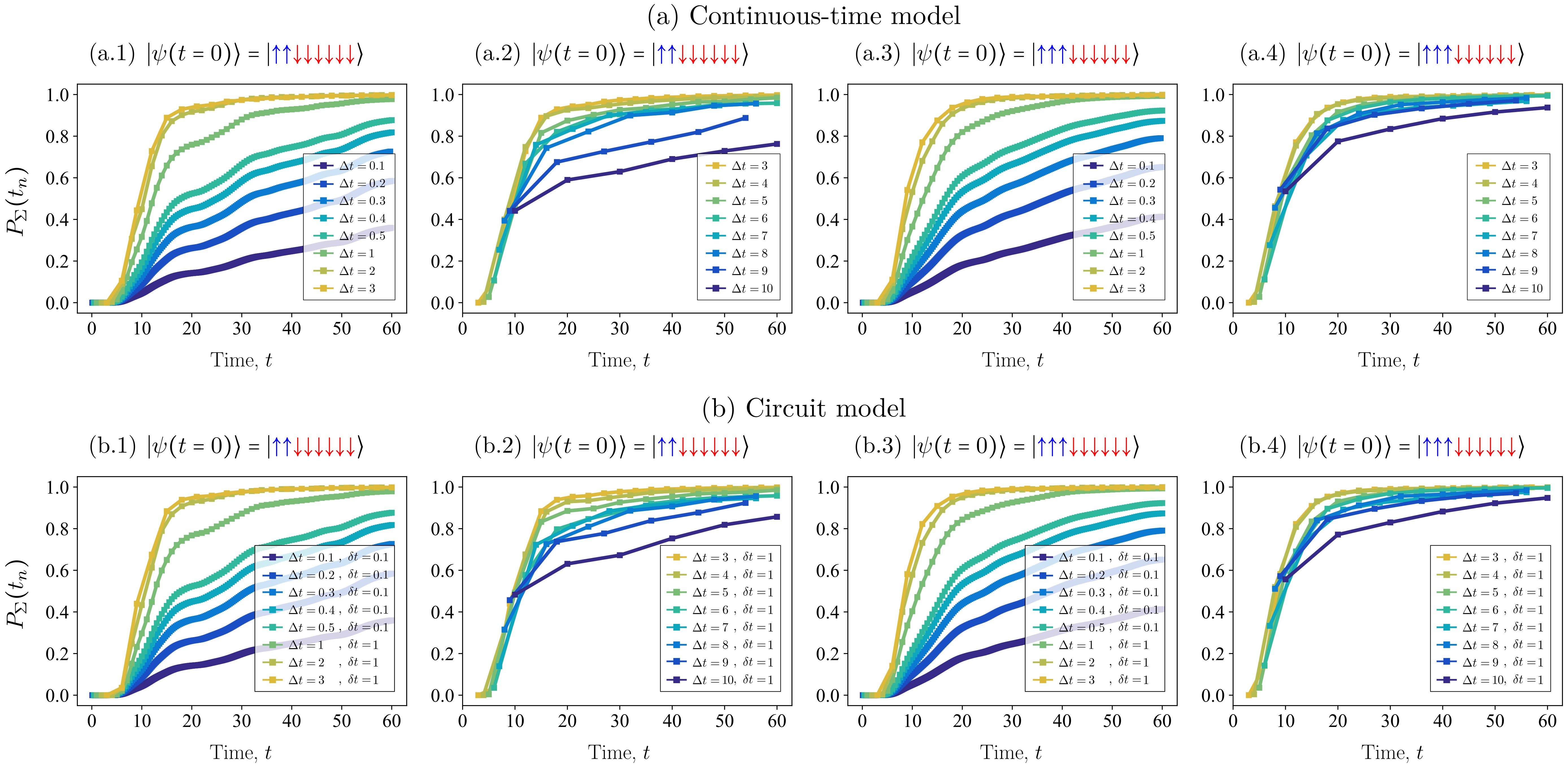}
    \caption{Arrival-time problem with multiple interacting particles in the different monitoring regimes.
(a): Cumulative arrival probability $P_\Sigma(t_n)$ in the continuous-time
model with anisotropy $\Delta=1/2$. (a.1): Two up-spins ($L=8$) from the
quantum Zeno regime up to the optimal monitoring period $\Delta t=3$.
(a.2): Same as (a.1) but in the low-monitoring regime, $\Delta t\geq3$.
(a.3) and (a.4): Same as (a.1) and (a.2) but for three up-spins ($L=9$).
(b): Same as (a) but for the circuit model. The Trotter step is
$\delta t=0.1$ for $\Delta t\leq0.5$ and $\delta t=1$ otherwise, and the
time is defined as $t=N\delta t$, where $N$ is the number of brickwork cycles.}
    \label{fig:7}
\end{figure*}

The late-time behavior is controlled by the spectrum of $\mathcal M$. Let
$\mu_1,\mu_2,\dots$ be its eigenvalues ordered by modulus, and define the
single-particle decay rates $\gamma_j=-2\ln|\mu_j|/\Delta t$. At late times
$\det[A_n^\dagger A_n]\propto\prod_{j\le k}|\mu_j|^{2n}$, so the survival
probability for generic monitoring periods and initial states is determined by the $k$ slowest rates as
\begin{equation}\label{eq:rates}
  P_s(t)\propto e^{-\gamma_{(k)}t},\qquad
  \gamma_{(k)}=\sum_{j=1}^{k}\gamma_j .
\end{equation}
Each additional up-spin adds its own decay channel, which explains why arrival
is faster for larger $k$ (see Fig.~\ref{fig:7} for details).

Without monitoring, the up-spin probability at the detector site is simply a sum
of single-particle contributions,
$\langle\hat P_\uparrow\rangle_L(t)=\sum_{j=1}^{k}|u_{Lj}(t)|^2$. The first
arrival is therefore set by the up-spin closest to the detector. A single
particle propagates with maximal velocity $J$ (see below), so its signal
reaches the detector after a time of order $(L-k)/J$.

\subsection{Optimal monitoring period}

\textit{Single-particle spectrum.}
For a single particle with amplitude $\psi_j$ on site $j$, the eigenvalue
equation in the bulk reads
\begin{equation}
  E\,\psi_j=\frac{J}{2}\big(\psi_{j+1}+\psi_{j-1}\big).
\end{equation}
A plane wave $\psi_j=e^{iqj}$ gives the band $E(q)=J\cos q$, where
$q\in(-\pi,\pi]$ is the quasi-momentum. The bandwidth is $W=2J$, and wave
packets move with group velocity $v(q)=dE/dq=-J\sin q$, at most $J$. In the
open chain the allowed momenta are $q_n=n\pi/(L+1)$, $n=1,\dots,L$. The
optimal monitoring period follows from comparing short and long periods.

\textit{Short periods ($J\Delta t\ll1$): quantum Zeno regime.}
Let $P=\hat P_\uparrow$ act on site $L$, and let $|\psi\rangle$ be the state
right after a no-click outcome, so that $P|\psi\rangle=0$. The click
probability at the next measurement is
$p_\uparrow=\Vert P\,U(\Delta t)|\psi\rangle\Vert^2$. Writing
$P\,U(\Delta t)=U(\Delta t)P(\Delta t)$ with the Heisenberg-picture projector
$P(s)=e^{iHs}Pe^{-iHs}$, and using $P|\psi\rangle=0$, we find
\begin{equation}
  P\,U(\Delta t)|\psi\rangle=U(\Delta t)\big[P(\Delta t)-P\big]|\psi\rangle .
\end{equation}
Since $dP(s)/ds=i\,e^{iHs}[H,P]e^{-iHs}$, we have
$\Vert P(\Delta t)-P\Vert\le\Delta t\,\Vert[H,P]\Vert$. Only the hopping
across the last bond changes the occupation of site $L$, so
\begin{equation}
  [H,P]=\frac{J}{2}\big(c_{L-1}^\dagger c_L-c_L^\dagger c_{L-1}\big),
  \qquad \Vert[H,P]\Vert=\frac{J}{2},
\end{equation}
and therefore
\begin{equation}\label{eq:zeno}
  p_\uparrow\le \frac{J^2\Delta t^2}{4}
\end{equation}
for every state, every $\Delta$, and every number of up-spins. By
Eq.~\eqref{eq:arrivalprob} the survival probability is a product of
no-click probabilities, each at least $1-J^2\Delta t^2/4$, so
\begin{equation}
  P_s(t_n)\ge\Big(1-\frac{J^2\Delta t^2}{4}\Big)^{n}\simeq
  e^{-J^2\Delta t\,t_n/4},
\end{equation}
and Eq.~(\ref{eq:arrival-time}) gives
\begin{equation}
  \langle t\rangle\ \ge\ \frac{4}{J^2\Delta t}.
\end{equation}
The arrival rate is thus at most $J^2\Delta t/4$ and vanishes as
$\Delta t\to0$. This is the quantum Zeno effect: the amplitude to reach the
detector grows only linearly in time, so each measurement finds the
detector empty with probability $1-O(\Delta t^2)$.


\textit{Long periods ($J\Delta t\gtrsim1$): dark states and the low-monitoring regime.}
If $\mathcal M$ has an eigenvalue with $|\mu|=1$, the corresponding
orbital passes every measurement without loss and is never detected.
Such a dark orbital exists if and only if $u(\Delta t)$ has an
eigenvector with no weight on site $L$. Since every energy eigenmode
of the open chain has weight on site $L$, this requires two
single-particle eigenphases to coincide,
\begin{equation}
  (E_a-E_b)\,\Delta t=2\pi\ell,\qquad \ell=1,2,\dots,
  \label{eq:bands}
\end{equation}
where $E_a>E_b$ \cite{Friedman2017PRE, Thiel2020}. In this case, the combination
$\langle L|E_b\rangle|E_a\rangle-
 \langle L|E_a\rangle|E_b\rangle$
cancels on the detector site at every measurement time. Physically,
the detector samples the dynamics periodically and cannot distinguish
phases that differ by multiples of $2\pi$. These resonances occur at
discrete monitoring periods. Close to such a period, an eigenvalue
of $\mathcal M$ approaches the unit circle, producing a slowly
decaying mode that can delay detection if it contributes to the
initial state's evolution. For several fermions, a single dark
orbital does not by itself prevent first detection: a completely
dark $k$-particle component requires at least $k$ independent dark
orbitals and nonzero initial overlap with their exterior-product
subspace. Stroboscopic resonances can therefore produce strongly
non-monotonic dependence on $\Delta t$, as observed in the low-monitoring regime, with their effect depending
on the initial state and particle number.


\textit{Optimal monitoring.}
The dark-state condition~\eqref{eq:bands} identifies a natural scale
for the monitoring period. For the finite open chain, the spectral
width is $W_L=2J\cos[\pi/(L+1)]$, so no single-particle dark orbital
can occur for $\Delta t<2\pi/W_L$. The first eigenphase degeneracy
appears at this threshold, although its effect on detection depends
on the initial-state overlap with the corresponding dark subspace.
Numerically, the minimum of the mean first detection time lies near
this scale for the systems considered in Fig.~\ref{fig:6}(a), suggesting
the estimate
\begin{equation}\label{eq:dtopt}
  \Delta t_{\rm opt}\approx\frac{2\pi}{W}
  =\frac{\pi}{J},
  \qquad W=2J,
\end{equation}
which gives $\Delta t_{\rm opt}\approx\pi$ for $J=1$.
This proximity is consistent with a balance between alleviating
Zeno suppression as the monitoring period increases and avoiding
slow decay associated with stroboscopic resonances. The bandwidth
therefore provides a physically motivated estimate of the optimal
period, rather than fixing its precise value. The weak dependence
of the observed optimum on particle number further suggests that
this single-particle spectral scale remains relevant to
many-particle first detection in the cases studied.

At first glance, there is no direct connection between the monitoring period that minimizes the mean first detection time and the one that maximizes the cumulative arrival probability
$P_\Sigma(t)$ at each time point. In the low-monitoring regime, the cumulative arrival curves for different monitoring periods can cross, so a faster buildup of arrival probability at early times can be accompanied by slower detection at later times. However, Fig.~\ref{fig:7} shows that, among the simulated monitoring periods, the curve near the optimal monitoring
period lies above those for the other periods throughout the displayed time window. This observation supports the agreement
between the two optimization criteria in our simulations: if such dominance holds for all times, it also guarantees a minimum of the
mean first detection time. Thus, we conclude that the optimal monitoring rate for $P_\Sigma(t)$ can also be obtained from \eqref{eq:dtopt}.

\subsection{Beyond the free-fermion model: $\Delta\neq0$}
For $\Delta\neq0$, the Jordan--Wigner transformation
\cite{JordanWigner1928} gives
\begin{equation}
    H=\sum_{ij}h_{ij}\,c_i^\dagger c_j
    +\frac{J\Delta}{2}(n_1+n_L)
    +J\Delta\sum_{j=1}^{L-1}n_jn_{j+1}
   +\text{const.},
\end{equation}
where terms proportional to the conserved particle number have
been absorbed into the constant within each fixed-$k$ sector.
The first correction is a one-body edge potential: an isolated
up-spin at an end site has one antiparallel bond instead of two,
giving an energy shift $J\Delta/2$ relative to the bulk.
The second correction is a nearest-neighbor interaction between
fermions, which generally breaks the free-fermion description.
If only the edge potential is retained, the Hamiltonian remains
quadratic, and Eqs.~\eqref{eq:D1}--\eqref{eq:arrival-time} remain
exact with
$h\rightarrow h+\frac{J\Delta}{2}(|1\rangle\langle1|+|L\rangle\langle L|)$.

\textit{Single up-spin.}
For $k=1$, the density interaction vanishes, so the anisotropy acts
only through the edge potential and the quadratic description
remains exact. The source and detector sites are shifted in energy
relative to the bulk, modifying the particle's escape from the
source and its subsequent detection. For the parameters considered
here, this increases the mean first detection time from
$\langle t\rangle\approx10.6$ at $\Delta=0$ to $\approx12.8$
at $\Delta=1/2$ for $\Delta t=3$
[Fig.~\ref{fig:6}(b)]. In a semi-infinite chain, an edge potential
of magnitude $|J\Delta|/2>J/2$, or $|\Delta|>1$, supports an
edge-localized state. In a finite chain, localization near the source can substantially delay detection, although the states associated with the two ends can hybridize.
The edge potential leaves the bulk dispersion unchanged but modifies the finite-chain eigenenergies and eigenvectors, so the optimal monitoring period remains fixed modulo these effects. For $\Delta=1/2$ our numerical results place the optimum near $\Delta t\approx\pi$, as in the free case.


\textit{Several up-spins.}
For $k\ge2$, the interaction $J\Delta\,n_jn_{j+1}$ allows
bound-magnon components in appropriate momentum sectors.
Magnon bound states were predicted by Bethe
\cite{bethe1931theorie}, identified in local-quench dynamics
\cite{ganahl2012observation}, observed in cold-atom spin chains
\cite{fukuhara2013microscopic}, and realized in XXZ circuits on
superconducting qubits \cite{morvan2022formation}.
Our initial states place the up-spins on adjacent sites and can
therefore overlap with these bound components, which coexist with
unbound scattering components. The cumulative arrival-time probabilities for multiple interacting spins are illustrated in Fig.~\ref{fig:7}. The slower propagation of bound components can delay detection, while the fast unbound component can retain a ballistic arrival onset similar to that of the noninteracting system. For the parameters considered here, the mean first detection time at $\Delta t=3$ increases from $\approx10.0$ to $\approx12.8$ for $k=2$ and from $\approx9.0$
to $\approx11.5$ for $k=3$ when the anisotropy changes from $\Delta=0$ to $\Delta=1/2$. A comparison with the model retaining only the edge potential gives a smaller increase, indicating that the density interaction contributes substantially to the slowdown.

Interactions also modify the stroboscopic resonances. A
single-particle dark orbital is generally no longer sufficient to characterize the interacting dynamics: a many-body dark state must evolve entirely within the subspace with an empty detector site at every measurement. Interactions can lift particular
free-fermion resonances, although they do not exclude many-body dark states in general. For the parameters and monitoring periods shown in Fig.~\ref{fig:6}(b), the several-particle curves exhibit broad resonances rather than the divergences seen in the single-particle case.

\section{Quantum simulation details}
\label{app:hardware}

\subsection{General considerations}

The quantum simulations presented in Figs.~\ref{fig3} and~\ref{fig4} were
executed on two of IBM's superconducting quantum processors,
\texttt{ibm\_kingston} and \texttt{ibm\_berlin}, accessed through the IBM
Quantum cloud platform~\cite{IBMQuantum}. Both devices are built from
fixed-frequency transmon qubits with tunable couplers, and both use CZ as the
native two-qubit gate together with the single-qubit gate set
$\{X,\sqrt{X},R_Z,R_X\}$, where $R_Z$ is implemented virtually.
\texttt{ibm\_kingston} is a 156-qubit Heron r2 processor whose qubits are
arranged in the heavy-hex lattice architecture, while \texttt{ibm\_berlin} is
a 120-qubit Nighthawk processor in which the qubits form a square lattice, so
that each qubit is connected to up to four nearest neighbors. This
square-lattice connectivity is naturally suited to the implementation of
two-dimensional models, such as the 2D model considered here, since the
interaction graph can be mapped onto the device without introducing
additional SWAP layers.

At the time of execution, \texttt{ibm\_kingston} exhibited median coherence
times of $T_1\approx250\,\mu\mathrm{s}$ and $T_2\approx140\,\mu\mathrm{s}$, a median
single-qubit gate error of $3.353\times10^{-4}$, a median CZ gate error of
$2.055\times10^{-3}$, and a median readout error of $8.73\times10^{-3}$. For
\texttt{ibm\_berlin} the corresponding median values were
$T_1\approx340\,\mu\mathrm{s}$, $T_2\approx235\,\mu\mathrm{s}$, a single-qubit gate
error of $1.844\times10^{-4}$, a CZ gate error of $2.024\times10^{-3}$, and a
readout error of $1.84\times10^{-2}$. Both processors therefore operate with
single- and two-qubit gate fidelities above $99.9\%$ and $99.7\%$,
respectively.

For benchmarking and validation purposes, we employed IBM's high-performance
simulators available through Qiskit~\cite{Qiskit2024}. These simulators
emulate IBM Quantum devices by incorporating noise models that account for
native gate errors, qubit connectivity, measurement errors, and decoherence.
We used the Sampler primitive~\cite{Qiskit2024}, which executes quantum
circuits repeatedly to produce bit-string distributions, capturing raw
hardware noise without any post-processing. Each circuit was executed with
20{,}000 measurement shots.

\subsection{Error mitigation}
\label{app:postselection}

To mitigate the effects of noise and decoherence, we applied two readily
available techniques: matrix-free measurement mitigation
(M3)~\cite{Nation2021M3} and dynamical decoupling
(DD)~\cite{ViolaLloyd1998,ViolaKnillLloyd1999,DuanGuo1999}. M3 operates
within a reduced subspace defined by the noisy input bit-strings, making the
associated linear system much easier to solve because the number of unique
bit-strings is far smaller than the full Hilbert-space dimension. DD, on the
other hand, helps preserve quantum coherence during idle periods by applying
carefully designed pulse sequences that dynamically refocus the qubit states.
In our circuits DD is mainly relevant during the mid-circuit measurements,
where the qubits that are not being read out idle for a comparatively long
time and would otherwise accumulate dephasing errors.

To improve the quality of the results obtained on the IBM devices, we exploit
the $U(1)$ symmetry of the dynamics, which conserves the total particle
number (equivalently, the total magnetization $\sum_i Z_i$), in the spirit of
symmetry-verification error mitigation~\cite{BonetMonroig2018,McArdle2019}.
Since neither the ideal unitary dynamics nor the $Z$-basis monitoring can
connect sectors with different particle number, any measured bitstring whose
Hamming weight differs from that of the initial state must originate from
hardware noise.

For each evolution time $t$, we execute the corresponding circuit and measure
all qubits in the computational basis $\{|0\rangle,|1\rangle\}$. We then
discard every shot (bitstring) that does not reproduce the particle number of
the input state, and estimate observables from the retained subset only. This
post-selection removes a substantial fraction of the corrupted counts, and requires no calibration data or noise model.
However, the retained fraction decreases with circuit depth, so deeper
circuits require a correspondingly larger number of shots to maintain the
same statistical uncertainty.

\subsection{Statistical confidence of the quantum simulation results}
\label{app:statistics}

\begin{figure}[t]
    \centering
    \includegraphics[width=1\linewidth]{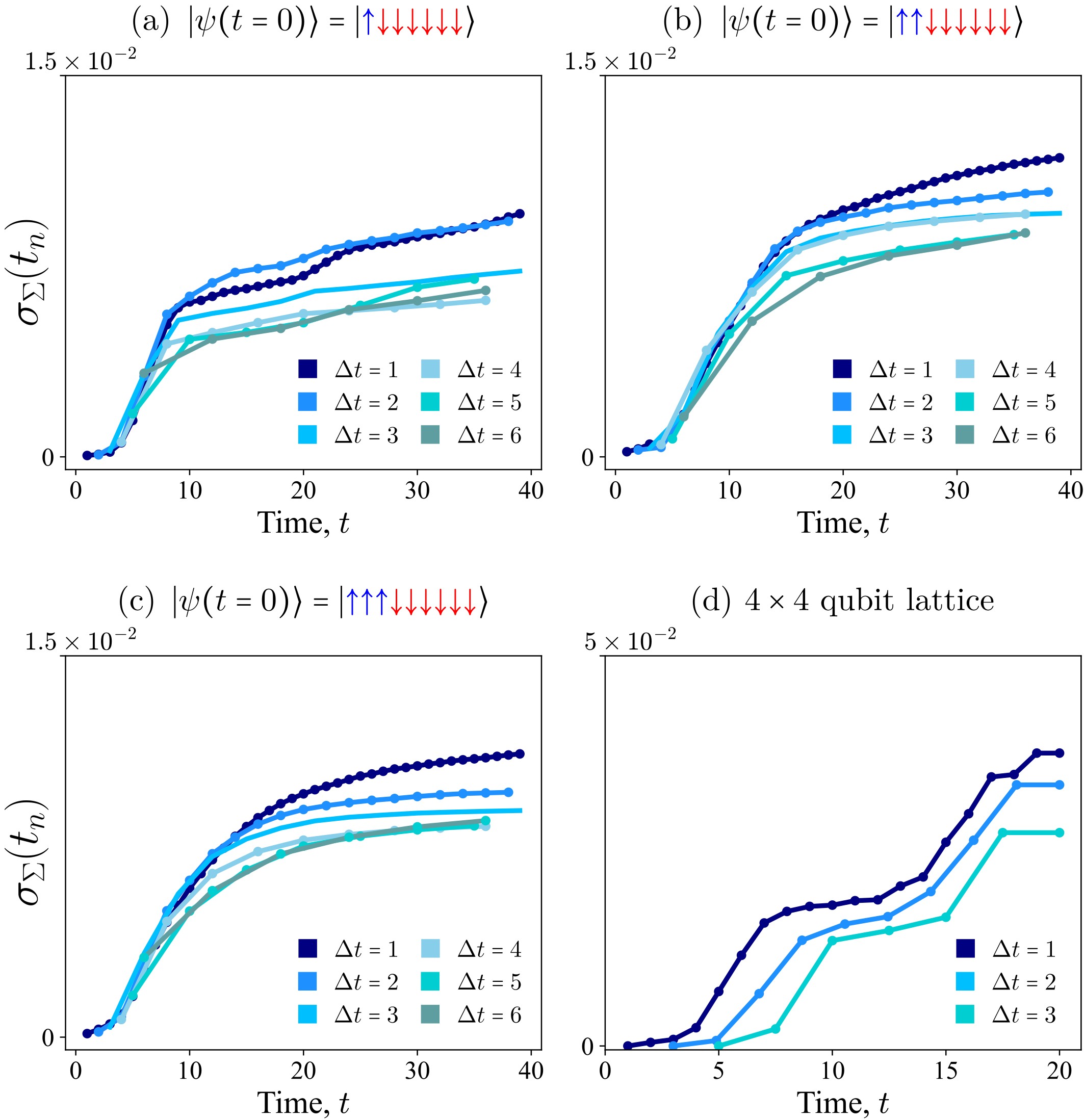}
    \caption{Statistical uncertainty of the measured arrival-time distributions.
Standard deviation $\sigma_\Sigma(t_n)$ of the cumulative arrival probability,
Eqs.~(\ref{eq:C1}) and (\ref{eq:C2}), obtained from $N_{\rm shot}=20{,}000$ shots per monitoring
time after the symmetry-based post-selection.
(a): Single up-spin initial state on \texttt{ibm\_kingston}, for the
monitoring periods of Fig.~\ref{fig3}(d).
(b) and (c): Same as (a) but for two and three up-spins, corresponding to
Figs.~\ref{fig3}(e) and (f).
(d): Same as (a) but for the $4\times4$ qubit lattice, corresponding to Fig.~\ref{fig4}(c).
In all cases $\sigma_\Sigma(t_n)$ grows monotonically with time, since the
variances of independent monitoring times add, but remains below
$1.5\times10^{-2}$ in 1d and $5\times10^{-3}$ in 2d, well below the
features of the arrival-time distributions.}
    \label{fig:std}
\end{figure}

The arrival probabilities at different monitoring times are obtained from
independent experiments. For each monitoring time $t_i$ we execute a separate
circuit with $N_{\mathrm{shot}}=20{,}000$ shots. After the symmetry-based
post-selection, $N_i$ shots are retained, of which $m_i$ exhibit the
monitoring record $(\downarrow,\dots,\downarrow,\uparrow)$ on the last qubit,
i.e., the first up-spin outcome occurs at $t_i$. The arrival probability is
estimated as $\hat P_a(t_i)=m_i/N_i$. Conditional on the number of retained
shots, $m_i$ is binomially distributed, giving
\begin{equation}
\sigma_a^2(t_i)=\frac{\hat P_a(t_i)\,[1-\hat P_a(t_i)]}{N_i}.
\label{eq:C1}
\end{equation}
The cumulative probability follows from Eq.~\eqref{eq:cumlativeprob}. Since the
individual terms are measured in independent circuits, their variances add,
\begin{equation}
\sigma_\Sigma^2(t_n)=\sum_{i=1}^{n}\sigma_a^2(t_i).
\label{eq:C2}
\end{equation}
Consequently, $\sigma_\Sigma(t_n)$ grows monotonically with $n$. Although the
retained fraction $N_i/N_{\mathrm{shot}}$ decreases with circuit depth, the
late-time arrival probabilities are small, so each late term contributes
little individually; their aggregate contribution is nevertheless
non-negligible due to the number of terms. Moreover, the cumulative estimates
at different times are correlated,
$\mathrm{Cov}[\hat P_\Sigma(t_n),\hat P_\Sigma(t_{n'})]
=\sigma_\Sigma^2(t_{\min(n,n')})$, since they share all terms up to the earlier time. A statistical fluctuation in an early term therefore shifts the entire subsequent curve, and the uncertainties quoted here are pointwise. In Fig.~\ref{fig:std}, we have plotted the standard deviation of the cumulative arrival probabilities obtained in the quantum simulation. This shows that the characteristic statistical error in the measured $P_\Sigma(t)$ is of the order of $1\%$ even at late times where it accumulates. This indicates that the accuracy of the experimental results is not limited by the statistical sample size but by intrinsic noise in the quantum processor.

\section{Monitored vs. unmonitored arrival time distributions}\label{app:unmonitored}

\begin{figure*}[t]
    \centering
    \includegraphics[width=1\linewidth]{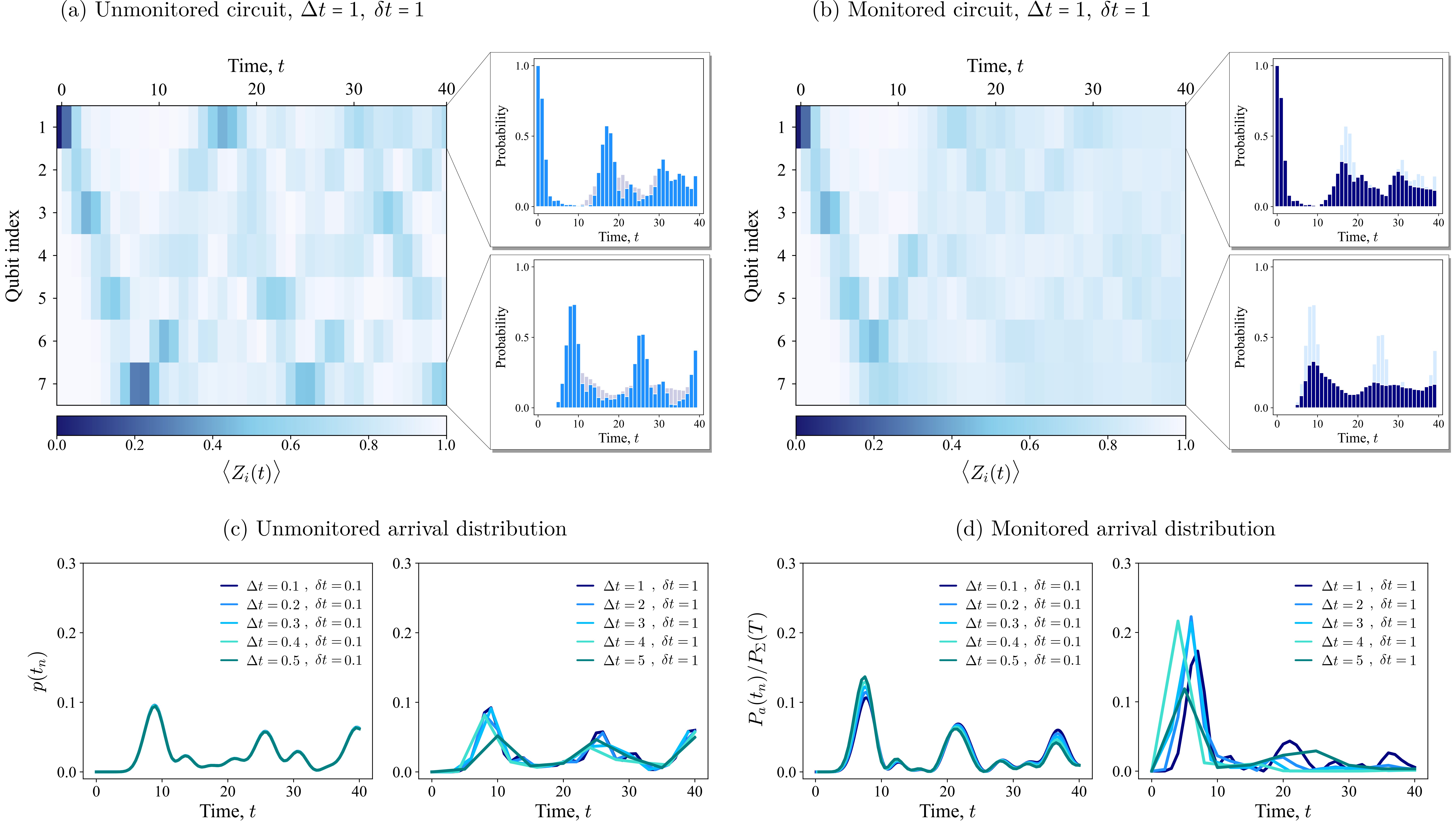}
 \caption{   
    Comparison of unmonitored and monitored dynamics for a single
up-spin initial state in the circuit model with $L=7$, $\Delta=1/2$ and
$\delta t=1$, in the representation of Ref.~\cite{Lloyd2025}.
(a): Local magnetization map of the unmonitored circuit (quantum stroboscopy). The insets show the up-spin
probability on the initial qubit (top) and on the monitored qubit (bottom).
(b): Same as (a) but with the last qubit monitored every $\Delta t=1$.
(c): Arrival distribution of the unmonitored circuit, $p(t_n)$, obtained by
normalizing the up-spin probability on the monitored qubit in the window
$T=40$. In the unmonitored case, $\Delta t$ only defines the sampling time $t_n=n\Delta t$ of the undisturbed evolution.  (d): Conditional first-detection probability
$P_a(t_n)/P_\Sigma(T)$ of the monitored circuit in the same window.}
    \label{fig:8}
\end{figure*}

To highlight the distinction between arrival-time prescriptions,
we compare our monitored first-detection distributions with the
recently introduced prescription based on independently evolved
copies of the system \cite{Lloyd2025}. The two protocols realize
operationally distinct temporal distributions
\cite{Martellini2025}, both of which can be accessed on the same
quantum hardware.

The protocol of Ref.~\cite{Lloyd2025}, termed quantum stroboscopy,
uses a freshly prepared copy of the system for each measurement,
with no monitoring before the chosen measurement time.
For a single particle, the measured probabilities
$w(x_i,t_n)$ of finding the particle at position $x_i$ and
measurement time $t_n$ are arranged into a matrix.
At each fixed time, these probabilities sum to one over position.
A temporal distribution at a fixed position is then obtained
by normalizing over the sampled times within a window $[0,T]$:
\begin{equation}
  p_{\rm st}(t_n\mid x_i;T)
  =\frac{w(x_i,t_n)}
  {\sum_{m:\,0\le t_m\le T}w(x_i,t_m)}.
\end{equation}
Here we assume equally spaced measurement times and equal
numbers of experimental shots at each time.

The required occupation probabilities can be obtained from
the local magnetization shown in Fig.~\ref{fig:8}(a), calculated
for an initial state with a single up-spin. The insets show
the up-spin probabilities on the first and last qubits before
normalization over time. For reference, Fig.~\ref{fig:8}(b)
shows the corresponding monitored dynamics.
Without monitoring, the excitation propagates through the chain
and is reflected at its ends, producing repeated peaks in the
occupation of the last site. For the parameters shown, the first
two prominent peaks occur near $t\sim L$ and $t\sim3L$, with
intervals of low occupation between them.
With monitoring every $\Delta t=1$, the evolution before
the first measurement is identical to the unmonitored evolution,
and the first-click probability equals the unmonitored detector
occupation at that time. Subsequent measurements modify the
dynamics through measurement backaction.

To compare the temporal profiles within a common window,
we interpret $p_{\rm st}(t_n\mid L;T)$ as the probability that
the chosen measurement time was $t_n$, conditioned on the fact that the particle is observed in the last site in the considered time window. Operationally, a measurement time
is selected uniformly from the sampling grid in $[0,T]$, and a
fresh copy is evolved to that time and measured. This
conditioning does not require the particle to have been absent
from the detector at earlier times; later returns also contribute.
For our monitored protocol, we instead use
\begin{equation}
  p_{\rm FD}(t_n\mid T)
  =\frac{P_a(t_n)}{P_\Sigma(T)},
  \qquad 0<t_n\le T,
\end{equation}
which is the probability of first detection at $t_n$,
conditioned on detection occurring within the same window.
These normalizations permit a comparison of the temporal
profiles, although the conditioning events are different.

Figures~\ref{fig:8}(c) and (d) compare the resulting distributions.
Both show a peak associated with the first propagation of the
excitation to the last site. However, the monitored
first-detection distribution depends strongly on the monitoring
period and can differ substantially from the quantum-stroboscopy
distribution. In the latter, the sampling interval determines
the temporal resolution, whereas in the monitored protocol it
also changes the physical evolution. Moreover, extending the
normalization window includes additional returns in the
unmonitored distribution and can change its normalized profile.
The comparison therefore illustrates how arrival-time
distributions depend on the measurement procedure and the
conditioning used to define them.

\bibliography{bibliography}
\end{document}